\documentclass[preprint,showpacs,showkeys,preprintnumbers,amsmath,amssymb]
{revtex4}

\usepackage{graphicx}
\usepackage{dcolumn}
\usepackage{bm}

\begin{document}

%\preprint{APS/123-QED}

\title{Two cavity modes in a dissipative environment: cross decay rates and
robust states}  

\author{R. Rossi Jr.}
\email{rossirj@fisica.ufmg.br}
\affiliation{Departamento de F\'{\i}sica, ICEX, Universidade Federal de
Minas Gerais, C.P. 702, 30161-970 Belo Horizonte, MG, Brazil}

\author{A. R. Bosco de Magalh\~aes}
\email{arbm@fisica.ufmg.br}
\affiliation{Departamento de F\'{\i}sica, ICEX, Universidade Federal de
Minas Gerais, C.P. 702, 30161-970 Belo Horizonte, MG, Brazil}

\author{M. C. Nemes}
\email{carolina@fisica.ufmg.br}
\affiliation{Departamento de F\'{\i}sica, ICEX, Universidade Federal
de Minas Gerais, C.P. 702, 30161-970 Belo Horizonte, MG, Brazil}

\date{\today}

\begin{abstract}
We investigate the role of the cross decay rates in a system composed by two
electromagnetic modes interacting with the same reservoir. Two feasible
experiments sensitive to the magnitudes and phases of these rates are
described. We show that if the cross decay rates are appreciable there are
states less exposed to decoherence and dissipation, and in limit situations
a decoherence free subspace appears.
\end{abstract}



\maketitle

It has become increasingly important to understand how the environment acts
over a system we are interested in. The promise of Quantum Computation often
has as its main obstacle the entanglement of the variables of interest with
the reservoir degrees of freedom \cite{art1}. Also, in the investigation of
Foundations of Quantum Mechanics, the environment plays a central role 
\cite{art2}.

A very successful model for the environment is the Caldeira-Legget model,
where one harmonic oscillator linearly coupled to a bath of oscillators is
considered \cite{art3}. 
The investigation of bipartite quantum systems interacting
with the environment usually displays amazing effects \cite{art4, art5}. 
In \cite{art6}, the
Caldeira-Legget model was extended for two oscillators subjected to the same
bath, leading to a possible communication between the oscillators mediated
by the reservoir. This extension may be used to study two electromagnetic
modes constructed in one or two cavities.

In \cite{art7}, 
an experiment dealing with two modes constructed in a single cavity
was described. The experimental findings were compared to theoretical
results in \cite{art8}, with satisfactory agreement. However, due to the large
detuning of the modes, the experiment is not sensitive to the cross decay
rates: the parameters related to the their communication through the
environment.

Cross decay rates and their interference effects over quantum systems are
well-known for a long time \cite{art9}. 
Some results related to them are population
trapping \cite{art10, art5} and decoherence free subspaces (DFS) 
\cite{art11, art12}. The
extension of the Caldeira-Legget model for two oscillators predicts the
possibility of DFS if the cross decay rates are large enough \cite{art8}. 
Given the
growing interest in DFS, the knowledge about this topic is accordingly
important. Thus, we propose, here, feasible experiments to investigate these
rates in the cavity Quantum Electrodynamics context, and also identify states
that are more resistent against decoherence and dissipation when the cross
decay rates are not zero.

The Hamiltonian we use to model two electromagnetic modes subjected to the
same environment is 
\begin{eqnarray}
\mathbf{H} &=&\mathbf{H}_{0}+\mathbf{H}_{int},  \label{H} \\
\mathbf{H}_{0} &=&\hbar \Omega _{1}\mathbf{a}_{1}^{\dagger }\mathbf{a}%
_{1}+\hbar \Omega _{2}\mathbf{a}_{2}^{\dagger }\mathbf{a}_{2}+\hbar 
\sum_{k} \omega _{k}\mathbf{c}_{k}^{\dagger }\mathbf{c}_{k}, 
\nonumber \\
\mathbf{H}_{int} &=&\hbar \sum_{k} \left( \alpha _{1k}\mathbf{a}%
_{1}\mathbf{c}_{k}^{\dagger }+\alpha _{1k}^{\ast }\mathbf{a}_{1}^{\dagger }%
\mathbf{c}_{k}\right) +\hbar \sum_{k} \left( \alpha _{2k}\mathbf{%
a}_{2}\mathbf{c}_{k}^{\dagger }+\alpha _{2k}^{\ast }\mathbf{a}_{2}^{\dagger }%
\mathbf{c}_{k}\right) ,  \nonumber
\end{eqnarray}
where $\mathbf{a}_{1}$, $\mathbf{a}_{2}$, $\mathbf{a}_{1}^{\dagger }$ and $%
\mathbf{a}_{2}^{\dagger }$ are annihilation and creation bosonic operators
for the modes of interest, $M_{1}$ and $M_{2}$, with frequencies $\Omega
_{1} $ and $\Omega _{2}$. The annihilation and creation operators $\mathbf{c}%
_{k}$ and $\mathbf{c}_{k}^{\dagger }$ are used to model the environment by a
set of harmonic oscillators linearly coupled to $M_{1}$ and $M_{2}$ modes.
This Hamiltonian shall be used for two modes in different cavities or in the
same cavity. Under the usual approximations, Hamiltonian (\ref{H}) leads to
the master equation 
\begin{equation}
\frac{d}{dt}\mathbf{\rho }_{S}\left( t\right) =\mathcal{L}\mathbf{\rho }%
_{S}\left( t\right) ,  \label{ME}
\end{equation}
with the Liouvillian superoperator (an operator which acts on operators) 
\begin{eqnarray}
\mathcal{L} &=&k_{11}\left( 2\mathbf{a}_{1}\bullet \mathbf{a}_{1}^{\mathbf{%
\dagger }}-\bullet \mathbf{a}_{1}^{\mathbf{\dagger }}\mathbf{a}_{1}-\mathbf{a%
}_{1}^{\mathbf{\dagger }}\mathbf{a}_{1}\bullet \right) +i\left( \Delta
_{11}-\Omega _{1}\right) \left[ \mathbf{a}_{1}^{\mathbf{\dagger }}\mathbf{a}%
_{1},\bullet \right] \mathcal{+}  \label{L} \\
&&k_{22}\left( 2\mathbf{a}_{2}\bullet \mathbf{a}_{2}^{\mathbf{\dagger }%
}-\bullet \mathbf{a}_{2}^{\mathbf{\dagger }}\mathbf{a}_{2}-\mathbf{a}_{2}^{%
\mathbf{\dagger }}\mathbf{a}_{2}\bullet \right) +i\left( \Delta _{22}-\Omega
_{2}\right) \left[ \mathbf{a}_{2}^{\mathbf{\dagger }}\mathbf{a}_{2},\bullet %
\right] \mathcal{+}  \nonumber \\
&&k_{12}\left( \mathbf{a}_{1}\bullet \mathbf{a}_{2}^{\mathbf{\dagger }}+%
\mathbf{a}_{2}\bullet \mathbf{a}_{1}^{\mathbf{\dagger }}-\bullet \mathbf{a}%
_{2}^{\mathbf{\dagger }}\mathbf{a}_{1}-\mathbf{a}_{1}^{\mathbf{\dagger }}%
\mathbf{a}_{2}\bullet \right) +  \nonumber \\
&&k_{21}\left( \mathbf{a}_{2}\bullet \mathbf{a}_{1}^{\mathbf{\dagger }}+%
\mathbf{a}_{1}\bullet \mathbf{a}_{2}^{\mathbf{\dagger }}-\bullet \mathbf{a}%
_{1}^{\mathbf{\dagger }}\mathbf{a}_{2}-\mathbf{a}_{2}^{\mathbf{\dagger }}%
\mathbf{a}_{1}\bullet \right) +  \nonumber \\
&&i\left( \frac{\Delta _{12}-\Delta _{21}}{2}\right) \left( \mathbf{a}%
_{1}\bullet \mathbf{a}_{2}^{\mathbf{\dagger }}-\mathbf{a}_{2}\bullet \mathbf{%
a}_{1}^{\mathbf{\dagger }}-\bullet \mathbf{a}_{2}^{\mathbf{\dagger }}\mathbf{%
a}_{1}+\mathbf{a}_{1}^{\mathbf{\dagger }}\mathbf{a}_{2}\bullet \right) + 
\nonumber \\
&&i\left( \frac{\Delta _{21}-\Delta _{12}}{2}\right) \left( \mathbf{a}%
_{2}\bullet \mathbf{a}_{1}^{\mathbf{\dagger }}-\mathbf{a}_{1}\bullet \mathbf{%
a}_{2}^{\mathbf{\dagger }}-\bullet \mathbf{a}_{1}^{\mathbf{\dagger }}\mathbf{%
a}_{2}+\mathbf{a}_{2}^{\mathbf{\dagger }}\mathbf{a}_{1}\bullet \right) + 
\nonumber \\
&&i\left( \frac{\Delta _{12}+\Delta _{21}}{2}\right) \left[ \mathbf{a}_{1}^{%
\mathbf{\dagger }}\mathbf{a}_{2}+\mathbf{a}_{2}^{\mathbf{\dagger }}\mathbf{a}%
_{1},\bullet \right] ,  \nonumber
\end{eqnarray}
for zero temperature, where 
\begin{equation}
k_{ij}+i\Delta _{ij}=\sum_{k}\alpha _{ik}\alpha _{jk}^{\ast }{\int_{0}^{\tau
_{c}}}d\tau e^{i\left( \omega _{k}-\Omega _{j}\right) \tau },
\label{k_ij+delta_ij}
\end{equation}
and we used the conventional notation for superoperators \cite{art13} 
(the dot sign (%
$\bullet $) indicates the place to be occupied by $\mathbf{\rho }_{S}\left(
t\right) $, where the superoperator acts). A discussion about the
Hamiltonian (\ref{H}) and a detailed derivation of the master equation (\ref
{ME}) from (\ref{H}) may be found in \cite{art6}. The coupling between the
environmental modes and the modes of interest may occur by complicated
processes, \emph{e. g.}, a photon may be scattered from a mode of interest
to an environmental mode by an atom. Of course, $H_{int}$ does not come from
a first principles analysis, and, thus, we do not know much about the
coupling constants $\alpha _{1k}$ and $\alpha _{2k}$. However, as we have
said, this model presents results that agree with experimental ones. The
cross decay rates are $k_{12}+i\Delta _{12}$ and $k_{21}+i\Delta _{21}$, and
will assume relevant values if the summations involving $\alpha _{1k}\alpha
_{2k}^{\ast }$ and $\alpha _{2k}\alpha _{1k}^{\ast }$ are not null. This
demands that the $\alpha _{1k}$ and the $\alpha _{2k}$ must have some
correlation, \emph{i. e.}, the way the modes of interest interact with the
environment must be microscopically correlated. This may be achieved if the
modes of interest are close to each other in the scale of the wavelength of
the environmental modes that effectively interacts with them. These are the
ones with frequencies around $\Omega _{1}$ or $\Omega _{2}$, as may be seen
in Eq. (\ref{k_ij+delta_ij}). If the environmental modes are electromagnetic
modes, the modes of interest must be close in the scale of their proper
wavelengths. An interesting case is two modes with orthogonal polarizations
constructed in the same cavity. They occupy essentially the same positions
in space, but maybe the difference in the polarizations spoils the
correlation.

\begin{figure}[!ht]
\includegraphics[width=16cm,height=13cm]{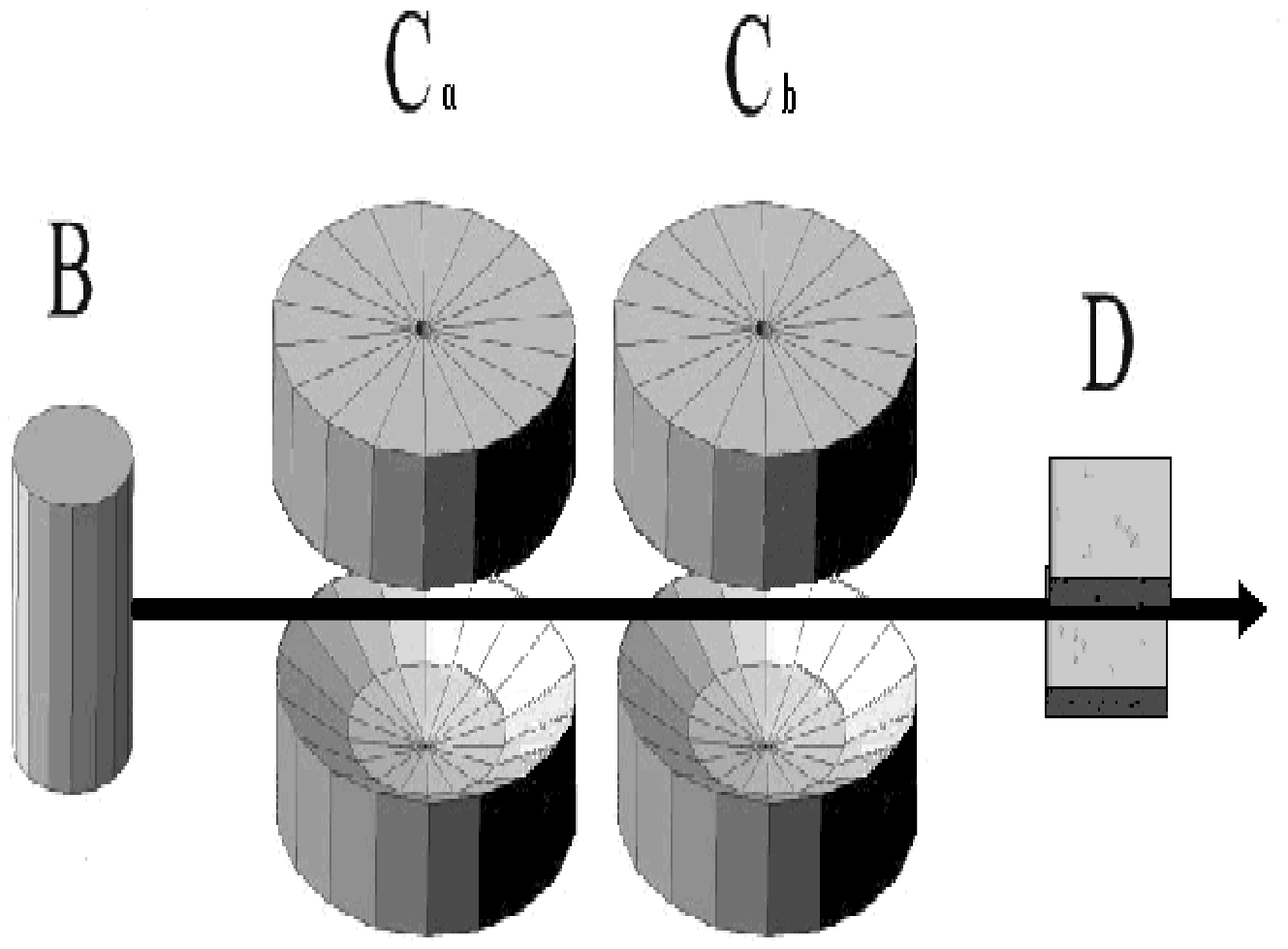}
\vspace{-4cm}
\caption{Sketch of the experiment with two cavities.}
\label{Experiment}
\end{figure}

In Fig. \ref{Experiment}, we sketch an experiment where circular Rydberg
atoms $\mathbf{A}_{\mathbf{s}}$ and $\mathbf{A}_{\mathbf{p}}$, with relevant
levels $e$ and $g$, are produced in box $\mathbf{B}$, cross the cavities $%
\mathbf{C}_{\mathbf{1}}$ and $\mathbf{C}_{\mathbf{2}}$ and are detected in
detector $\mathbf{D}$. The energy of level $e$ is higher than the energy of
level $g$ by $\hbar \Omega _{a}$. The cavities are macroscopically
identical: $\Omega _{1}=\Omega _{2}=\Omega $ and $k_{11}=k_{22}=k$. We will
assume a huge number of environmental modes, with random distribution in the
frequencies around $\Omega $, what leads to $\Delta _{11}=\Delta _{22}\simeq
0$ and $k_{12}+i\Delta _{12}\simeq \left( k_{21}+i\Delta _{21}\right) ^{\ast
}$, as may be seen using expression (\ref{k_ij+delta_ij}). We may adjust the
atom-fields interaction time using the Stark effect: when an appropriate
voltage is applied across the mirrors of the cavities, the $e\leftrightarrow
g$ transition becomes resonant with the modes $M_{1}$ and $M_{2}$; when no
voltage is applied, the detuning $\Delta =\Omega _{a}-\Omega $ turns
negligible the atom-fields interaction.

In the analysis below, global phases will be often disregarded. The cavities
are initially in vacuum state, and $\mathbf{A}_{\mathbf{s}}$ is sent in
state $e$: 
\[
\left| \psi \left( t=0\right) \right\rangle _{M_{1},M_{2},\mathbf{A}_{%
\mathbf{s}}}=\left| 0,0,e\right\rangle . 
\]
Atom $\mathbf{A}_{\mathbf{s}}$ enters cavity $\mathbf{C}_{\mathbf{1}}$, and
is put in resonance with mode $M_{1}$ by a time $t_{1s}$, yielding 
\[
\left| \psi \left( t=t_{1s}\right) \right\rangle _{M_{1},M_{2},\mathbf{A}_{%
\mathbf{s}}}=\cos \left( Gt_{1s}\right) \left| 0,0,e\right\rangle -i\sin
\left( Gt_{1s}\right) \left| 1,0,g\right\rangle , 
\]
where we used the RWA (rotating wave approximation) Hamiltonian 
\[
\mathbf{H}_{1s}=\hbar \Omega \left( \mathbf{a}_{1}^{\dagger }\mathbf{a}_{1}+%
\mathbf{a}_{2}^{\dagger }\mathbf{a}_{2}+\frac{\sigma _{z}}{2}\right) +\hbar
G\left( \mathbf{a}_{1}^{\dagger }\sigma _{-}+\mathbf{a}_{1}\sigma
_{+}\right) , 
\]
with $\sigma _{z}=\left| e\right\rangle \left\langle e\right| -\left|
g\right\rangle \left\langle g\right| $, $\sigma _{-}=\left| g\right\rangle
\left\langle e\right| $, $\sigma _{+}=\left| e\right\rangle \left\langle
g\right| $. During a time $t_{0s}$, $\mathbf{A}_{\mathbf{s}}$ is put far of
resonance with the modes. Considering $\left| \Delta \right| \gg G$, the
Hamiltonian may be written 
\[
\mathbf{H}_{0s}=\hbar \Omega \left( \mathbf{a}_{1}^{\dagger }\mathbf{a}_{1}+%
\mathbf{a}_{2}^{\dagger }\mathbf{a}_{2}\right) +\frac{\hbar \Omega
_{a}\sigma _{z}}{2}, 
\]
and the system evolves to 
\[
\left| \psi \left( t=t_{1s}+t_{0s}\right) \right\rangle _{M_{1},M_{2},%
\mathbf{A}_{\mathbf{s}}}=\cos \left( Gt_{1s}\right) \left|
0,0,e\right\rangle -ie^{i\left( \Omega _{a}-\Omega \right) t_{0s}}\sin
\left( Gt_{1s}\right) \left| 1,0,g\right\rangle . 
\]
Next, $\mathbf{A}_{\mathbf{s}}$ enters $\mathbf{C}_{\mathbf{2}}$, and is put
in resonance with $M_{2}$ by $t_{2s}=\pi /\left( 2G\right) $. Considering
that $M_{1}$ and $M_{2}$ have the same polarizations, the Hamiltonian will
be 
\[
\mathbf{H}_{2s}=\hbar \Omega \left( \mathbf{a}_{1}^{\dagger }\mathbf{a}_{1}+%
\mathbf{a}_{2}^{\dagger }\mathbf{a}_{2}+\frac{\sigma _{z}}{2}\right) +\hbar
G\left( \mathbf{a}_{2}^{\dagger }\sigma _{-}+\mathbf{a}_{2}\sigma
_{+}\right) , 
\]
what leads to 
\begin{equation}
\left| \psi \left( t=t_{1s}+t_{0s}+t_{2s}\right) \right\rangle
_{M_{1},M_{2}}=\cos \theta \left| 0,1\right\rangle +e^{i\phi }\sin \theta
\left| 1,0\right\rangle ,  \label{S1}
\end{equation}
where $\theta =Gt_{1s}$ and $\phi =\left( \Omega _{a}-\Omega \right) t_{0s}$%
. The $\mathbf{A}_{\mathbf{s}}$ state ends up in $g$, factorized, and needs
not to be considered anymore.

During the time $T$ ($T\gg t_{1s}+t_{0s}+t_{2s}$), no atom interacts with
the modes. Using Eq. (\ref{ME})\ to compute the action of the environment in
this period, we get, for the state of $M_{1}\leftrightarrow M_{2}$ system, 
\begin{eqnarray}
\mathbf{\rho }_{S}\left( t=T+t_{1s}+t_{0s}+t_{2s}\right) &=&\left\{
u_{1}\left( T\right) \left| 1,0\right\rangle +u_{2}\left( T\right) \left|
0,1\right\rangle \right\} \left\{ \text{HC}\right\}  \label{Evolucao1} \\
&&+\left| 0,0\right\rangle \left\langle 0,0\right| \left\{ 1-\left|
u_{1}\left( T\right) \right| ^{2}-\left| u_{2}\left( T\right) \right|
^{2}\right\} ,  \nonumber
\end{eqnarray}
where HC stands for Hermitian conjugate, and 
\begin{eqnarray*}
u_{1}\left( t\right) &=&\frac{e^{-kt}}{2}\left\{ \left(
e^{-rt}+e^{rt}\right) e^{i\phi }\sin \theta +re^{-i\gamma }\left(
e^{-rt}-e^{rt}\right) \cos \theta \right\} , \\
u_{2}\left( t\right) &=&\frac{e^{-kt}}{2}\left\{ \left(
e^{-rt}+e^{rt}\right) \cos \theta +re^{i\gamma }\left( e^{-rt}-e^{rt}\right)
e^{i\phi }\sin \theta \right\} , \\
re^{i\gamma }\text{\ \ } &=&k_{12}+i\Delta _{12}.
\end{eqnarray*}
In this calculation, we used the parameter differentiation technic,
explained in \cite{art6}. 
At time $t=t_{1s}+t_{0s}+t_{2s}+T$, atom $\mathbf{A}_{%
\mathbf{p}}$, initially in state $g$, starts to interact with mode $M_{1}$.
Considering that $\mathbf{A}_{\mathbf{p}}$ interacts with $M_{1}$ during a
time $t_{1p}=3\pi /\left( 2G\right) $, with $M_{2}$ during $t_{2p}=\left(
2\theta -\pi \right) /\left( 2G\right) $, and a time $t_{0p}=$ $t_{0s}$
between these interactions, the probability to find the atom $\mathbf{A}_{%
\mathbf{p}}$ in state $e$ at $t=t_{1s}+t_{0s}+t_{2s}+T+t_{1p}+t_{0p}+t_{2p}$
will be 
\[
P_{e}=\frac{e^{-2kT}}{4}\left| \left( e^{-rT}+e^{rT}\right) +\sin 2\theta
\cos \left( \gamma +\phi \right) \left( e^{-rT}-e^{rT}\right) \right| ^{2}. 
\]
This is the main result of the present contribution. Observe that $P_{e}=1$
if there is no environment, since in this case $k=r=0$.

\begin{figure}[!ht]
\includegraphics[width=16cm,height=15cm]{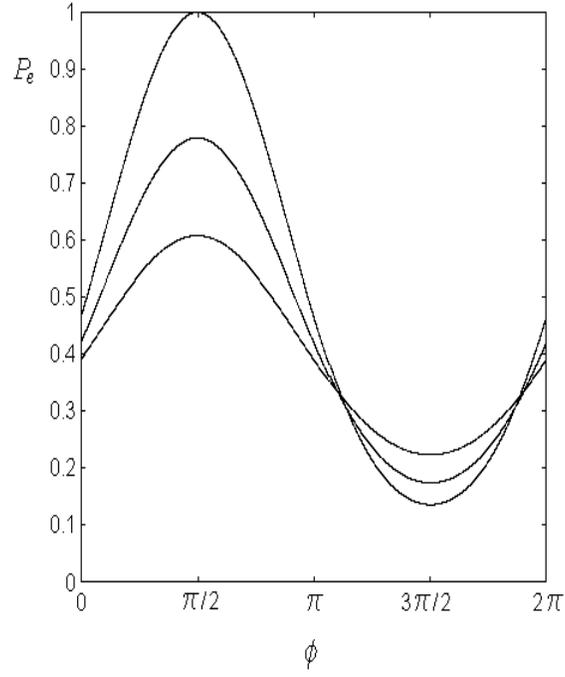}
\vspace{-3cm}
\caption{Probability $P_{e}$ for $%
k=1000$ $s^{-1}$, $\protect\theta =\protect\pi /4$, $\protect\gamma =\protect%
\pi /2$, $T=500$ $\protect\mu s$ and various values for $r$: $r=500$ $s^{-1}$%
, $r=750$ $s^{-1}$ and $r=1000$ $s^{-1}$. Higher $r$ values correspond to
higher amplitudes in the oscillation of the curve.}
\label{plot2c}
\end{figure}

The probability $P_{e}$ depends on $\theta $ and $\phi $, which may be
freely chosen by varying the times $t_{1s}$, $t_{0s}$ and $t_{2s}$. For
fixed $T$ and $\theta $, a $P_{e}\times \phi $ plot constructed with
experimental data tells us about the cross decay rates: if any sinusoidal
pattern is observed, $k_{12}+i\Delta _{12}$ is non zero, the amplitude of
the curve being related to $r$; the position of the maximum of the curve may
be used to find $\gamma $ (this maximum occurs for $\gamma +\phi =\pi $).
This is exemplified in Fig. \ref{plot2c}. The larger visibility is achieved
for $\theta =\pi /4$, \emph{i. e.}, when the state (\ref{S1}) is maximally
entangled. Since the magnitudes of the cross decay rates are related to
correlations between the way each mode interacts with the environment, we
expect that the amplitudes of the curves grow when the cavities get closer.

In order to get better insight into how the robust states arise in the model,
let us rewrite Eq. (\ref{L}) in the form 
\begin{eqnarray*}
\mathcal{L} &=&\mathcal{L}_{1}\mathcal{+L}_{2}, \\
\mathcal{L}_{1} &=&\left( k-r\right) \left( 2\mathbf{A}_{1}\bullet \mathbf{A}%
_{1}^{\dagger }-\bullet \mathbf{A}_{1}^{\dagger }\mathbf{A}_{1}-\mathbf{A}%
_{1}^{\dagger }\mathbf{A}_{1}\bullet \right) +i\Omega \left[ \mathbf{A}%
_{1}^{\dagger }\mathbf{A}_{1},\bullet \right] , \\
\mathcal{L}_{2} &=&\left( k+r\right) \left( 2\mathbf{A}_{2}\bullet \mathbf{A}%
_{2}^{\dagger }-\bullet \mathbf{A}_{2}^{\dagger }\mathbf{A}_{2}-\mathbf{A}%
_{2}^{\dagger }\mathbf{A}_{2}\bullet \right) +i\Omega \left[ \mathbf{A}%
_{2}^{\dagger }\mathbf{A}_{2},\bullet \right] ,
\end{eqnarray*}
where the operators $\mathbf{A}_{1}$ and $\mathbf{A}_{2}$, given by 
\[
\left( 
\begin{array}{c}
\mathbf{A}_{1} \\ 
\mathbf{A}_{2}
\end{array}
\right) =\frac{1}{\sqrt{2}}\left( 
\begin{array}{cc}
1 & -\exp \left( -i\gamma \right) \\ 
\exp \left( i\gamma \right) & 1
\end{array}
\right) \left( 
\begin{array}{c}
\mathbf{a}_{1} \\ 
\mathbf{a}_{2}
\end{array}
\right) , 
\]
obey the usual commutation relations for bosons. Notice that
the decay rate related to $\mathcal{L}_{1}$ is smaller than the one related
to $\mathcal{L}_{2}$. Thus, states which may be written as 
\begin{equation}
\mathbf{\rho }_{S}=\sum_{m,n} c_{m,n}\left( \mathbf{A}%
_{1}^{\dagger }\right) ^{n}\left| 0,0\right\rangle \left\langle 0,0\right| 
\mathbf{A}_{1}^{m}+\text{HC}  \label{SLEE}
\end{equation}
are less exposed to the environment. When $r=k$, these states are
decoherence-free, a probably very difficult condition to reach. If we learn
the value of $\gamma $, by performing the experiment just described, we may
chose to work with the states less exposed to the environment (\ref{SLEE}).
Relevant examples for Quantum Information and Quantum Optics are the
maximally entangled state 
\[
\mathbf{\rho }_{S}=\left( \left| 1,0\right\rangle -e^{i\gamma }\left|
0,1\right\rangle \right) \left( \left\langle 1,0\right| -e^{-i\gamma
}\left\langle 0,1\right| \right) 
\]
and the coherent state 
\[
\mathbf{\rho }_{S}=\left| v,-e^{i\gamma }v\right\rangle \left\langle
v,-e^{i\gamma }v\right| . 
\]

As discussed above, the physical distance between the cavities may lead to
the destruction of the microscopic correlations important for the appearing 
of DFS and robust states. There is, however, another possibility, which
circumvents this problem: one can use two modes differing by their 
polarization in a single cavity. Is this a better proposition than the 
first one?
In order to investigate this, let us consider two degenerate modes
with orthogonal polarizations and equal dissipation rates. A
circular Rydberg atom with $e\leftrightarrow g$ transition resonant with the
modes will interact simultaneously with them via the Hamiltonian 
\[
\mathbf{H}_{12a}=\hbar \Omega \left( \mathbf{a}_{1}^{\dagger }\mathbf{a}_{1}+%
\mathbf{a}_{2}^{\dagger }\mathbf{a}_{2}+\frac{\sigma _{z}}{2}\right) +\hbar
G\left( i\mathbf{a}_{1}^{\dagger }\sigma _{-}-i\mathbf{a}_{1}\sigma _{+}+%
\mathbf{a}_{2}^{\dagger }\sigma _{-}+\mathbf{a}_{2}\sigma _{+}\right) . 
\]
The phase difference in the coupling constants of the atom to each mode are
due to the polarization orthogonality. Consider that one atom in state $e$
begins to interact with the modes, initially in vacuum state, at time $t=0$,
and this interaction lasts until the time $t_{12a}=\pi /(2\sqrt{2}G)$. This
atom creates the entangled state 
\begin{equation}
\mathbf{\rho }_{S}\left( t=t_{12a}\right) =\frac{1}{\sqrt{2}}\left( \left|
0,1\right\rangle +i\left| 1,0\right\rangle \right) \left( \text{HC}\right)
\label{ros1}
\end{equation}
for the modes in the cavity, and ends up in state $g$. Taking into account
the action of the environment in the period between $t=t_{12a}$ and $%
t=t_{12a}+T$ ( $T\gg \pi /(2\sqrt{2}G)$), the state of the fields will be
given by Eq. (\ref{Evolucao1}) with $u_{1}\left( t\right) $ and $u_{2}\left(
t\right) $ calculated using $\theta =\pi /4$ and $\phi =\pi /2$. Then, if
another atom, initially in state $g$, starts its interaction with the modes
at time $t=t_{12a}+T$ , and this interaction lasts until $t=2t_{12a}+T$, we
get 
\[
P_{e,r}=\frac{e^{-2kT}}{4}\left| \left( e^{-rT}+e^{rT}\right) -\sin \left(
\gamma \right) \left( e^{-rT}-e^{rT}\right) \right| ^{2} 
\]
for the probability to find this second atom in state $e$. Notice that $%
P_{e}=1$ if we disregard the environment ($k=r=0$). Another way to have $%
P_{e}=1$ is $k=r$ and $\gamma =\pi /2$: in this case we have a DFS.

If the cavity is subjected to a pressure, in such a way that it becomes
slightly elliptical, the orthogonal modes become non resonant. For a large
enough detuning, we may work with the Hamiltonian 
\[
\mathbf{H}_{12a}=\hbar \Omega \left( \mathbf{a}_{2}^{\dagger }\mathbf{a}_{2}+%
\frac{\sigma _{z}}{2}\right) +\hbar G\left( \mathbf{a}_{2}^{\dagger }\sigma
_{-}+\mathbf{a}_{2}\sigma _{+}\right) , 
\]
related to an atom interacting with just one resonant mode. Now, an atom,
initially in state $e$, interacts with the field mode, initially in vacuum
state, between $t=0$ and $t=\pi /2G$, yielding the field state $\left|
1\right\rangle $ and ending up in state $g$. If we take into account the
action of the environment between $t=\pi /2G$ and $t=\pi /2G+T$ ($T$ $\gg
\pi /2G$), and let a second atom, initially in $g$ state, to interact with
the field in the period between $t=\pi /2G+T$ and $t=\pi /G+T$, we get 
\[
P_{e,nr}=e^{-2kT} 
\]
for the probability to find this atom in state $e$.

\begin{figure}[!ht]
\includegraphics[width=16cm,height=15cm]{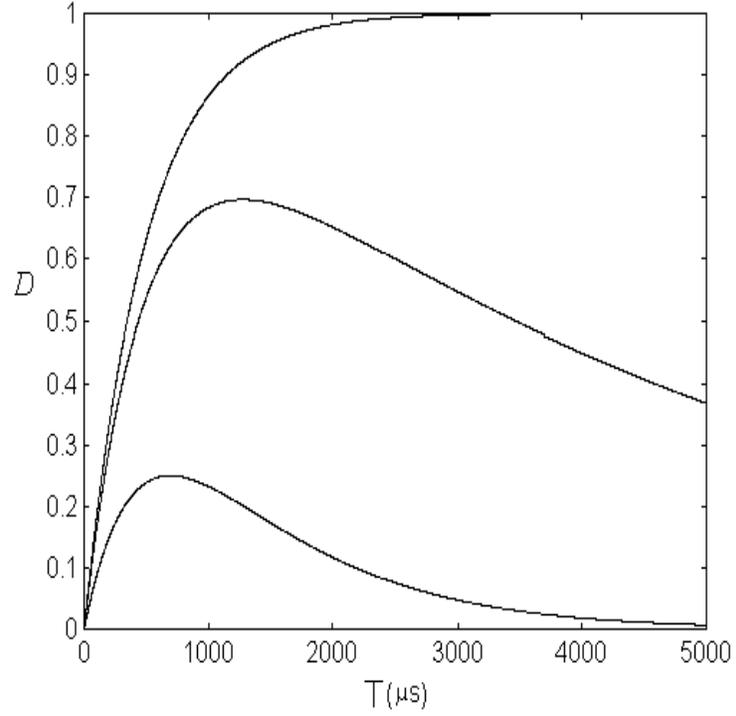}
\vspace{-3cm}
\caption{Difference between 
probabilities $D=P_{e,r}-P_{e,nr}$ for $k=1000$ $s^{-1}$, $\protect\gamma =%
\protect\pi /2$ and various values for $r$: $r=500$ $s^{-1}$, $r=900$ $%
s^{-1} $ and $r=1000$ $s^{-1}$. Higher $r$ values correspond to higher
curves.}
\label{plot1c}
\end{figure}

The difference between probabilities $D=P_{e,r}-P_{e,nr}$ must be a sign of
significative cross decay rates. In the experiment with two cavities, we can
maximize the effects of the parameter $r$ by choosing the state built by the
first atom; in this experiment with a single cavity, the first atom always
constructs the state (\ref{ros1}), and the effects of $r$ will be maximized
for $\gamma =\pi /2$ only. This is the condition that makes $D$ more easy to
detect, and it was used in Fig. \ref{plot1c}. Since $\gamma $ depends on the
environment, it is not trivial to control it.

Cross decay is related to interference: if the environment acts over both
modes in a microscopic correlated way, this action may be canceled (at least
partially) for a set of states. For the ideal case of perfect correlation,
the cross decay rates are large, and a DFS appears. If we get some
correlation (even non perfect), the knowledge about the cross decay rates
may be useful, since some states will be more resistent to the degradation
produced by the environment. The experiments we proposed here are feasible
with the present technological state, and the confirmation of these
interference effects in Cavity Quantum Electrodynamics systems could
encourage a search of analogous behavior in other contexts, maybe in
scalable systems, which would be important for Quantum Information
implementation.

The authors acknowledge financial support from the Brazilian agency
CNPq.

\end{document}